\documentstyle{amsart}
\newtheorem{theorem}{Theorem}[section]
\newtheorem{lemma}[theorem]{Lemma}
\newtheorem{corollary}[theorem]{Corollary}
\theoremstyle{definition}
\newtheorem{definition}[theorem]{Definition}
\newtheorem{definitions}[theorem]{Definitions}
\theoremstyle{remark}
\newtheorem{fact}{Fact}  
\newcommand{\pr}{\operatorname{pr}}
\newcommand{\lcm}{\operatorname{lcm}}

\begin{document}
\title{A Criterion for Stability in Random Boolean Cellular Automata}
\author{James F. Lynch}
\address{Department of Mathematics and Computer Science \\
Clarkson University \\
Potsdam, N. Y. 13699-5815}
\email{jlynch@@sun.mcs.clarkson.edu}
\thanks{Research supported by NSF Grant CCR-9006303.}
\maketitle
\begin{abstract}
Random boolean cellular automata are investigated, where each gate has
two randomly chosen inputs and is randomly assigned a boolean function
of its inputs. The effect of non-uniform distributions on the choice of
the boolean functions is considered. The main results are that if the
gates are more likely to be assigned constant functions than
non-canalyzing functions, then with very high probability, the automaton
will exhibit very stable behavior: most of the gates will stabilize, and
the state cyles will be bounded in size.
\end{abstract}
\section{Introduction}
Boolean cellular automata are models of parallel computation that have
attracted much attention from researchers in complex systems and
artificial life. Computer simulation of these automata have shown that they
often exhibit stable and robust behavior, even when randomly constructed.
The implications of this evidence have been described in numerous articles
by S. Kauffman and others (see for example \cite{k},
which includes an extensive bibliography). Only recently, however, have
rigorous mathematical methods been applied to the study of boolean
cellular automata. This article is a continuation of the efforts begun
in {\L}uczak and Cohen \cite{lc} and Lynch \cite{l}. We investigate
randomly constructed boolean cellular automata, where each gate has two
inputs, as in most of Kauffman's simulations. However, instead of
randomly assigning one of the 16 boolean functions of two arguments
to each gate with equal probability, we consider the effect of
non-uniform distributions, using a classification of boolean functions
introduced by Kauffman (op. cit.).
(We still require some mild symmetry conditions
on the probabilities.)

The boolean functions can be partitioned into the canalyzing and
noncanalyzing functions. A formal definition will be given in the next
section, but for now it suffices to note that among the canalyzing
functions are the constant functions; i.e. the function that outputs 0
regardless of its inputs and its negation that always outputs 1. Further,
among the two-argument boolean functions, there are only two non-canalyzing
functions: the {\sc equivalence} function that outputs 1 if and only if
both of
its inputs have the same value, and its negation the {\sc exclusive or}.

Our main result is that if the function assigned
to each gate is more likely to be constant than non-canalyzing, then with
very high probability the automaton will exhibit very stable behavior.
Specifically, it will have these four properties:
\begin{enumerate}
\item Almost all of the gates in the automaton will stabilize, regardless
of the starting state.
That is, they eventually settle into a state (0 or 1) that never changes.
\item Almost all of the gates are weak, that is, changing their state
does not affect the state cycle that is entered.
\item From any starting state, the state cycle will be entered quickly.
\item The state cycles are bounded in size.
\end{enumerate}
This shows, perhaps
surprisingly, that the nonconstant canalyzing functions, which include
the {\sc or} and the {\sc and} functions, have a neutral effect on the
stability of the automaton. It is the non-canalyzing gates that seem to
be the sources of instability.

Our results, and the earlier results in \cite{lc} and \cite{l} support
the belief that stability and emergent order are widespread
phenomena in boolean cellular automata. In addition, they give a simple,
exact condition that implies stability. Further, the distributions
where the
probabilities of constant and non-canalyzing gates are equal
(as in Kauffman's model) appear to
be thresholds between very stable and more complex behavior. This will be
described in a future article. At present, it is known to the author
that when the two probabilities are equal, with high probability almost
all of the gates still stabilize and are weak, but the state cycles
are no longer bounded in size. Most of them are larger than
$n^c$ for some $c > 0$. Very little is known about the behavior of random
automata when the probability
of non-canalyzing gates exceeds that of constant gates.

\section{Definitions}
We will now give precise definitions of the notions that were alluded to
in the previous section.
Let $n$ be a natural number. A {\em boolean cellular automaton}
with $n$ gates consists
of a directed graph $D$ with vertices $1,\dots,n$ (referred to as {\em
gates})
and a sequence ${\boldsymbol f} = (f_1,\dots,f_n)$ of boolean functions.
In this article, each gate will have indegree two, and each boolean
function will have two arguments.
We say that gate $j$ is an {\em input} to gate $i$ if
$(j,i)$ is an edge of $D$. A boolean cellular automaton $B=
\langle D,{\boldsymbol f}\rangle$ defines a map from $\{0,1\}^n$
(the set of 0-1
sequences of
length $n$) to $\{0,1\}^n$ in the following way. For each $i=1,\dots,n$ let
$j_i < k_i$ be the inputs of $i$.  Given $x =
(x_1,\dots,x_n) \in \{0,1\}^n$, $B(x) =
(f_1(x_{j_1},x_{k_1}),\dots,f_n(x_{j_n},x_{k_n}))$.
$B$ may be regarded as a finite state automaton with state set $\{0,1\}^n$
and initial state $x$. That is, its state at time 0 is $x$, and if its
state at time $t$ is $y \in \{0,1\}^n$ then its state at time $t+1$ is
$B(y)$. Our first set of definitions pertains to the aspects of
stability that will be studied.
\begin{definitions} Let $B$ be a boolean cellular automaton
and $x \in \{0,1\}^n$.
\begin{enumerate}
\item We put $B^t(x)$ for the state of $B$ at time $t$, and
$f_i^t(x)$ for the value of its $i$th component, or gate, at time $t$.
\item Since the number of states is finite, i.e. $2^n$, there exist times
$t_0 < t_1$ such that $B^{t_0}(x) = B^{t_1}(x)$. Let $t_1$ be the
first time at which this occurs. Then $B^{t+t_1-t_0}(x) = B^t$ for all
$t \geq t_0$. We refer to the set of states $\{B^t(x) : t \geq t_0 \}$
as the {\em state cycle\/} of $x$ in $\langle D,{\boldsymbol f} \rangle$, to
distinguish it from a cycle of $D$ in the graph-theoretic sense.
\item The
{\em tail\/} of $x$ in $\langle D,{\boldsymbol f}\rangle$ is $\{B^t(x) :
t < t_0
\}$.
\item Gate $i$ {\em stabilizes\/} in $t$ steps on input $x$
  if for all $t^\prime \ge
  t$, $f_i^{t^\prime}(x) = f_i^t(x)$.
\item Gate $i$ is {\em weak\/} if for any input $x$, letting $\overline{x}^i$
  be identical to $x$ except that its $i$th component is $1 - x_i$,
$$
{\boldsymbol \exists} t_0 {\boldsymbol \exists} d {\boldsymbol \forall} t
(t \ge t_0 \Rightarrow B^t(x) =
  B^{t + d}(\overline{x}^i)).
$$
That is, changing the state of $i$ does not affect the state cycle that
is entered.
\end{enumerate}
\end{definitions}

While we are primarily interested in stability, the related notion of
forcing seems to be easier to deal with combinatorially. Thus most of
our results pertain to forcing in boolean cellular automata, but as will
be evident, they translate directly into results about stability.
\begin{definitions}
Let $f(x_1,x_2)$ be a boolean function of two arguments.
\begin{enumerate}
\item We say that $f$
{\em depends\/} on argument $x_1$ if for some $v \in \{0,1\}$,
$f(0,v) \neq f(1,v)$. A symmetric definition applies when $f$ depends on
$x_2$. Similarly, if $\langle D,{\boldsymbol f}\rangle$ is a boolean
cellular automaton, $f_i = f$, and the inputs of gate $i$ are $j_{i1}$
and
$j_{i2}$, then for $m = 1,2$, $i$ depends on $j_{im}$ if $f$ depends on $x_m$.
\item
The function $f$ is said to be {\em canalyzing\/} if
there is some $m = 1$ or $2$ and some values $u,v \in \{0,1\}$ such that
for all $x_1,x_2 \in \{0,1\}$, if $x_m = u$ then $f(x_1,x_2) = v$.
Argument $x_m$ of $f$
is said to be a {\em forcing argument\/} with {\em forcing value\/} $u$ and
{\em forced value\/} $v$. Likewise, if $\langle D,{\boldsymbol f}\rangle$
is a boolean
cellular automaton and $f_i$ is a canalyzing function with forcing argument
$x_m$, forcing value $u$ and forced value $v$, then input $j_{im}$ is a {\em
forcing input\/} of gate $i$.  That is, if the value of $j_{im}$ is $u$ at time
$t$, then the value of $i$ is guaranteed to be $v$ at time $t+1$.
\end{enumerate}
\end{definitions}

All of these definitions generalize immediately to boolean
functions of arbitrarily many arguments. In the case of two argument
boolean functions, the only non-canalyzing functions are
{\sc equivalence} and {\sc exclusive or}. The two constant functions
$f(x,y) = 0$ and $f(x,y) = 1$
are trivially canalyzing, as are the four functions that depend on only
one argument:
\begin{align*}
f(x,y) & =  x, \\
f(x,y) & =  \neg x, \\
f(x,y) & =  y \text{, and}\\
f(x,y) & =  x.
\end{align*}
The remaining eight boolean functions of two arguments are canalyzing,
and they are all similar in the sense that both arguments are forcing with
a single value, and there is one forced value. A typical example is the
{\sc or} function. Both arguments are forcing with 1,
and the forced value is 1.

\begin{definition} Again, $\langle D,{\boldsymbol f}\rangle$
is a boolean cellular automaton.
Using induction on $t$, we define what it means for gate $i$ to be {\em forced
to a value\/} $v$ {\em in\/} $t$ {\em steps}.

If $f_i$ is the constant function $f(x_1,x_2)=v$, then $i$ is forced to $v$ in
$t$ steps for all $t \ge 0$.

If the inputs $j_{i1}$ and $j_{i2}$ of $i$ are forced to $u_1$ and $u_2$
respectively in $t$ steps, then $i$ is forced to $f_i(u_1,u_2)$
in $t+1$ steps.

If $f_i$ is a canalyzing function with forcing argument $x_m$,
forcing value $u$,
and forced value $v$, and $j_{im}$ is forced to $u$ in $t$ steps, then
$i$ is forced to $v$ in $t+1$ steps.
\end{definition}
By induction on $t$ it can be seen that if $i$ is forced in $t$ steps, then
it stabilizes for all initial states $x$ in $t$ steps.

The following combinatorial notions will be used in characterizing
forcing structures. We assume the reader is familiar with the basic
concepts of graph theory (see e.g. Harary \cite{h}). Unless otherwise
stated, {\em path\/} and {\em cycle\/} shall mean directed path and
cycle in the digraph $D$.
\begin{definitions}
\begin{enumerate}
\item For any gate $i$ in $\langle D,{\boldsymbol f}\rangle$ with inputs
$j_{i1}$ and $j_{i2}$, let
\begin{align*}
N_0^-(i) & =  \{i\} \text{ and} \\
N_{d+1}^-(i) & =  N_d^-(j_{i1}) \cup N_d^-(j_{i2}).
\end{align*}
\item Then
$$
S_d^-(i) = \bigcup_{c \leq d} N_c^-(i).
$$
That is, $S_d^-(i)$ is the set of all gates that are connected to $i$ by
a path of length at most $d$.
\item In a similar way we define $N_d^+(i)$ and $S_d^+(i)$, the set of all
gates reachable from $i$ by paths of length at most $d$.
\item For any nonnegative integer $d$, a $d$-{\em unforced path\/}
is a sequence of distinct gates
$P = (i_1,\ldots,i_p)$ such that $i_{r+1}$ depends on $i_r$ for
$1 \leq r < p$ and none of the gates are forced in
$d$ steps.
\item A $d$-{\em unforced
cycle\/} is the same except $i_1 = i_p$.
\end{enumerate}
\end{definitions}
Note that whether $i$ is forced in $d$ steps is completely determined by
the
restriction of $D$ and ${\boldsymbol f}$ to $S_d^-(i)$.

We will examine the asymptotic behavior of {\em random\/} boolean
cellular automata. For each boolean function $f$ of two arguments,
we associate a probability $a_f \in [0,1]$, where $\sum_f a_f = 1$.
The random boolean cellular automaton with $n$ gates is the result of
two random processes. First, a random digraph where every gate has indegree
two is generated. Independently for each gate, its two inputs are
selected from the \begin{math}
\bigl( \begin{smallmatrix}
n \\ 2
\end{smallmatrix} \bigr)
\end{math}
equally likely possibilities. Next, each gate is independently assigned a
boolean function of two arguments, using the probability distribution
$\langle a_f : f:\{0,1\}^2 \rightarrow \{0,1\}\rangle$.
We will use $\tilde{B} = \langle
\tilde{D},\tilde{\boldsymbol f}\rangle$ to denote a random boolean cellular
automaton generated as
above. For any properties $\cal P$ and $\cal Q$ pertaining to boolean cellular
automata, we put $\pr({\cal P},n)$ for the probability that the random boolean
cellular automaton on $n$ gates has property $\cal P$ and $\pr({\cal P}|
{\cal Q},n)$ for the conditional probability that $\cal P$ holds, given that
$\cal Q$ holds. Usually, we will omit the $n$ in these expressions since it
will be understood.

We classify the two argument boolean functions as follows:
\begin{enumerate}
\item ${\cal A}$ contains the two constant functions.
\item ${\cal B}_1$ contains the four canalyzing functions that depend on
one argument.
\item ${\cal B}_2$ contains the eight canalyzing functions that depend on
both arguments.
\item ${\cal C}$ contains the two non-canalyzing functions.
\end{enumerate}
Then the probabilities that a gate is assigned a function in each of the
categories are:
\begin{align*}
a & =  \sum_{f \in {\cal A}} a_f \\
b_1 & =  \sum_{f \in {\cal B}_1} a_f \\
b_2 & =  \sum_{f \in {\cal B}_2} a_f \\
c & =  \sum_{f \in {\cal C}} a_f
\end{align*}
Lastly, we put ${\cal B} = {\cal B}_1 \cup {\cal B}_2$ and
$b = b_1 + b_2$, the probability that a gate is assigned a nonconstant
canalyzing function. Throughout the rest of the article, we assume the
following symmetry conditions on our distributions:
\begin{align*}
a_f & = a_{\neg f} \text{ for all } f \in {\cal A} \cup {\cal B}_2, \\
a_{f(x,y)} & = a_{f(y,x)} \text{ for all } f \in {\cal B}_1.
\end{align*}
Also, $\log$ shall always mean $\log_2$.

\section{Stable Gates}
As previously mentioned, a gate is stable if it is forced. Thus, a lower
bound on the probability that a gate is forced also holds for the
probability of stability.
\begin{lemma}\label{lemforced}
For $d \ge 0$ and $v = 0,1$ let
\begin{align*}
p_d(v) & =  \pr(\text{gate $i$ is forced to $v$ in $d$ steps } \\
 &\qquad| S_d^-(i) \text{ induces an acyclic subgraph of } \tilde{D})
 \text{ and} \\
p_d & =  p_d(0) + p_d(1).
\end{align*}
Then \stepcounter{theorem}
\begin{equation}\label{eqprob}
p_d(0) = p_d(1)
\end{equation}
and $p_d$  satisfies the following recurrence:
\begin{align*}
p_0 & =  a, \\
p_{d+1} & =  a + b p_d + c p_d^2.
\end{align*}
\end{lemma}
\begin{pf}
We prove Equation (\ref{eqprob}) by induction on $d$.
The recurrence relation will be a byproduct.
For $d = 0$, it is clear since
$p_0(0) = p_0(1) = a/2$.

Next we prove Equation (\ref{eqprob}) for $d+1$, assuming it is true for $d$.
Let $j$ and $k$ be the two inputs of $i$. Since $S_{d+1}^-(i)$ induces
an acyclic subgraph, so do $S_d^-(j)$ and $S_d^-(k)$. The possible ways
that $i$ can be forced to $v$ are:
\begin{enumerate}
\item It is assigned the constant function $f(x,y) = v$.
\item It is assigned some function $f \in {\cal B}_1$, and the input on which
$i$ depends is forced in $d$ steps to the value that forces $f$ to $v$.
\item It is assigned some function $f \in {\cal B}_2$, $v$ is the forced
value of $f$
and at least one
of its inputs is forced in $d$ steps to the value that forces $f$ to $v$,
or $v$ is not the forced value of $f$ but $j$ and $k$ are forced to
values $u$ and $w$ such that $f(u,w) = v$.
\item It is assigned some function $f \in {\cal C}$, and $j$ and $k$ are
forced to values $u$ and $w$ such that $f(u,w) = v$.
\end{enumerate}
We will derive expressions for the probability of each of the four cases,
and show they are the same for $v = 0$ and 1. The probability of Case
(1) is $a/2$.

If $f \in {\cal B}_1$, say $f(x,y) = x$, the probability that $i$ is
forced to v in $d+1$ steps is $p_d(v)$. The other choices for $f \in
{\cal B}_1$ are symmetric, and by the induction assumption
the probability of Case (2) is $b_1p_d(0)$.

In Case (3), it can be observed that the eight functions in ${\cal B}_2$
may be partitioned into four pairs, each of the form $\{f,\neg f\}$.
Take a typical $f \in {\cal B}_2$, say the {\sc or} function. Then
0 is not the forced value of $f$, but it is for $\neg f$. The
probability that $f$ is forced to 0 is
$$
p_d(0)^2,
$$
and the probability
that $\neg f$ is forced to 0 is
$$
(1 - p_d(1))p_d(1) + p_d(1)(1 - p_d(1)) + p_d(1)^2.
$$
Summing these two probabilities and using symmetry and the induction
hypothesis, we
get the probability that the function assigned to $i$ is $f$ or $\neg f$
and $i$ is forced to 0 in $d+1$ steps:
$$
a_f \times p_d(0)^2 + a_{\neg f} \times(2p_d(1) - p_d(1)^2) =
  (a_f + a_{\neg f}) \times p_d(0).
$$
Summing over all four pairs of functions, the probability of Case (3) is
$b_2p_d(0)$. The argument when $v=1$ is symmetric.

Lastly, let $f \in {\cal C}$, say $f$ is {\sc exclusive or}. The probability
that $i$ is forced to 0 in $d+1$ steps is $p_d(0)^2 + p_d(1)^2$, and the
probability that $i$ is forced to 1 in $d+1$ steps is
$2p_d(0)p_d(1)$. By the induction assumption, these probabilities are
equal. Similar reasoning applies when $f$ is {\sc equivalence}. Thus
the probability of Case (4) is $2cp_d(0)^2$ regardless of whether $v$ is
0 or 1.

We have shown that for $v = 0$ or 1,
$$
p_{d+1}(v) = \frac{a}{2} + bp_d(0) + 2cp_d(0)^2,
$$
proving Equation (\ref{eqprob}).
Furthermore
$$
p_{d+1} = a + bp_d + cp_d^2. \qed
$$
\renewcommand{\qed}{}\end{pf}
\begin{corollary}\label{corstab}
We have $p_d > 1 - (1 - a + c)^{d+1}$.
\end{corollary}
\begin{pf}
Let $q_d = 1 - p_d$. By the Lemma,
\begin{align*}
q_0 & =  1 - a < 1 - a + c \text{ and} \\
1 - q_{d+1} & =  a + b(1 - q_d) + c(1 - q_d)^2.
\end{align*}
Since $a+b+c = 1$,
\begin{align*}
q_{d+1} & =  (1 - a + c)q_d - cq_d^2 \\
 & \le  (1 - a + c)q_d
\end{align*}
and the result follows by induction on $d$.
\end{pf}
\begin{theorem}\label{thmstable}
Let $a > c$.
For any positive $\alpha < 1/2$ there is a constant $\beta > 0$ such that
$$
\lim_{n \rightarrow \infty}\pr(\tilde{B} \text{ has at least
  $n(1 - n^{-\beta})$ gates that
  stabilize in $\alpha\log n$ steps}) = 1.
$$
\end{theorem}
\begin{pf}
We use the following slight modification of a Fact from \cite{lc}.
\begin{fact} For any positive $\alpha$ and function $\omega(n)$ that
increases to infinity,
$$
\lim_{n \rightarrow \infty}\pr(\tilde{B} \text{ has at most
$\omega(n)n^{2\alpha}$
  gates belonging to cycles of length $< \alpha\log n$}) = 1.
$$
\end{fact}

Let ${\bold Y}$ be the random variable that counts the number of gates in
$\tilde{B}$ that do not belong to a cycle of length $< \alpha\log n$
and do not stabilize in $\alpha\log n$ steps. Let ${\bold E}({\bold Y})$
be its expectation.
By Corollary \ref{corstab},
\begin{align*}
{\bold E}({\bold Y}) & \le  n(1 - a + c)^{\alpha\log n} \\
 & =  n^{1 - \gamma} \text{ for some $\gamma > 0$.}
\end{align*}
By Markov's inequality,
\begin{align*}
\pr({\bold Y}  \ge  n^{1 - \gamma/2}) & \le n^{-\gamma/2} \\
 & \rightarrow  0 \text{ as $n \rightarrow \infty$.}
\end{align*}
This implies, together with the Fact, that with probability asymptotic to
1, there are at most
$$
n^{1 - \gamma/2} + \omega(n)n^{2\alpha}
$$
gates that do not stabilize in $\alpha\log n$ steps, for any function
$\omega$ that increases to infinity. Taking $\beta = \min(\gamma/2,
1 - 2\alpha)$ and $\omega = \log$, the Theorem follows.
\end{pf}

\section{Unstable Structures}
We now study the sizes and shapes of the unstable components of $\tilde{B}$.
Actually, we will be looking at unforced components. Since a collection of
gates is unforced if it is unstable, showing that the unforced structures
have certain restrictions implies the same for unstable structures. The
next lemma is central to all these results.
\begin{lemma}\label{lemunforcedpath}
For any nonnegative integers $d$ and $l$ such that $l \le (\log n)^2$,
$$
\pr(\tilde{B} \text{ has a $d$-unforced path of length $l$}) \le
  n[2(1 - a + c)^d + b + 2c + o(1)]^l.
$$
\end{lemma}
\begin{pf}
We select a chain of $l+1$ gates as follows. Begin with $i_{l+1}$.
Then select
the two element set $I_{l+1}$ of inputs to $i_{l+1}$. From $I_{l+1}$,
select the gate
which is the predecessor of $i_{l+1}$ in the chain, and call it $i_l$.
Call the other gate $j_l$. Repeat this selection process with
$i_l$, and so on, ending with $i_1$ and $j_1$. The number of possible
sequences $(i_1,\dots,i_{l+1})$ and $(j_1,\dots,j_l)$ that can be
selected this way is bounded by
$$
n \times \left[
\bigl( \begin{smallmatrix}
n \\ 2
\end{smallmatrix} \bigr)
\times 2 \right]^l.
$$
Also,
$$
\pr(\bigwedge_{r = 1}^l ((i_r,i_{r+1}) \in \tilde{D} \wedge
  (j_r,i_{r+1}) \in \tilde{D})) =
  \bigl( \begin{smallmatrix}
n \\ 2
\end{smallmatrix} \bigr)^{-l}.
$$
For $r = 0,\dots,l+1$ let ${\cal P}_r$ be the event that
$(i_1,\dots,i_r)$ is a $d$-unforced path. We will finish the proof by
showing that
$$
\pr({\cal P}_r) \le [(1 - a + b)^d + (b + 2c)/2 + o(1)]^r.
$$
This will follow from
$$
\pr({\cal P}_r | {\cal P}_{r-1}) \le (1 - a + b)^d + (b + 2c)/2 + o(1).
$$
To prove this, let ${\cal Q}_r$ be the event that
$N_{d-1}^-(j_r)$ does not induce a tree and
$N_{d-1}^-(j_r)\cap \bigcup_{s = 1}^{r-1} N_d^-(i_s) = \emptyset$.
First, we show that \stepcounter{theorem}
\begin{equation}\label{eqnotree}
\pr(\neg {\cal Q}_r) = o(1).
\end{equation}

Now ${\cal Q}_r$ fails
only if there exists a path P of length $p \le d$
beginning at some gate $k$ and ending at $i_r$
and another path $Q$ of length $q$, $1 \le q \le d$, beginning at $k$,
disjoint from $P$ except at $k$ and possibly its other endpoint,
which must be in
$P$ or $\{i_1,\dots,i_r\}$. There are no more than $n^p$
ways of choosing $P$ and no more than $n^{q-1}\times(p+r)$ ways of choosing
$Q$. The probability of any such choice is bounded above by
$(2/n)^{p+q}$. Therefore the probability that $P$ and $Q$ exist is bounded
above by
$$
\sum_{p = 0}^d \sum_{q = 1}^d 2^{p+q}(p+r)n^{-1} = O((\log n)^2 n^{-1}),
$$
proving Equation (\ref{eqnotree}).

Now we
examine the conditional probability of ${\cal P}_r$, given that ${\cal P}_{r-1}
\wedge {\cal Q}_r$.
One possibility is that $j_{r-1}$ is not forced in $d-1$ steps.
Since ${\cal Q}_r$ holds, this event is independent of ${\cal P}_{r-1}$,
and by
Corollary \ref{corstab} this has probability $\le (1 - a + c)^d$. The
other possibility is that $j_{r-1}$ is forced in $d-1$ steps, but
$i_r$ is not forced in $d$ steps. There are three cases to consider:
\begin{enumerate}
\item $f_{i_r} \in {\cal B}_1$
\item $f_{i_r} \in {\cal B}_2$
\item $f_{i_r} \in {\cal C}$
\end{enumerate}
In Case (1), the input on which $i_r$ depends must be $i_{r-1}$ and not
$j_{r-1}$. Given that $f_{i_r} \in {\cal B}_1$, the probability that
$i_r$ depends on $i_{r-1}$ is $1/2$ because of the symmetry condition
$a_{f(x,y)} = a_{f(y,x)}$. Thus the probability of Case (1) is $b_1/2$.

In Case (2), $f_{i_r}$ can be forced by a single value on either input.
Since $j_{r-1}$ is forced, it must be forced to the value $v$ that does not
force $f_{i_r}$. Given that $j_{r-1}$ is forced, by Lemma \ref{lemforced},
the conditional probability that it is forced to $v$ is $1/2$. Therefore
the probability of Case (2) is $b_2/2$.

The probability of Case (3) is c.
Altogether,
$$
\pr({\cal P}_r | {\cal P}_{r-1}) \le \pr(\neg {\cal Q}_r) +
(1 - a + c)^d + \frac{b}{2} + c,
$$
and the Lemma follows.
\end{pf}
\begin{lemma}\label{lemunforcedcycle}
For any nonnegative integers $d$ and $l$ such that $l \le (\log n)^2$,
$$
\pr(\tilde{B} \text{ has a $d$-unforced cycle of length $l$}) \le
  [2(1 - a + c)^d + b + 2c + o(1)]^l.
$$
\end{lemma}
\begin{pf}
The same proof as in Lemma \ref{lemunforcedpath} applies, except the factor
$n$ disappears because
$i_1 = i_{l+1}$.
\end{pf}
\begin{lemma}\label{lemunforcedstart}
For any nonnegative integers $d$ and $l$ such that $l \le (\log n)^2$,
for any gate $i$ in $\tilde{B}$,
\begin{multline*}
\pr(\tilde{B} \text{ has a $d$-unforced path of length $l$ beginning
  at $i$}) \\
  \le [2(1 - a + c)^d + b + 2c + o(1)]^l.
\end{multline*}
\end{lemma}
\begin{pf}
The same proof applies here because $i_1 = i$ is already chosen.
\end{pf}
\begin{theorem}
If $a > c$ then there exists $\alpha > 0$ such that
$$
\lim_{n \rightarrow \infty}\pr(\tilde{B} \text{ has at least
  $n(1 - n^{-\alpha})$ weak gates }) = 1.
$$
\end{theorem}
\begin{pf}
Choose $d$ so that $2(1 - a + c)^d + b + 2c < 1$. This is possible because
$a > c$ and $a + b + c =1$. Take $l = \log n/3$ and let ${\bold Y}$ be the
random variable that counts the number of gates in $\tilde{B}$ that do not
belong to a cycle of length $< l$ and are not weak. Consider any such
gate $i$. If all $d$-unforced chains starting at $i$ are of length $< l$,
then the gates in $N_l^+(i)$ are not affected by the state of $i$.
Thus the state of $i$
does not affect the state cycle that $\tilde{B}$ enters, and $i$ would be
weak. Therefore there must be a $d$-unforced path beginning at $i$ of
length $l$. By Lemma \ref{lemunforcedstart},
$$
{\bold E}({\bold Y}) \le n^{1 - \gamma} \text{ for some $\gamma > 0$.}
$$
The rest of the proof proceeds as in the proof of
Theorem \ref{thmstable}.
\end{pf}
\begin{lemma}\label{lempathbound}
Let $a > c$. For sufficiently large $d$ there exists a constant $\beta$
such that
$$
\lim_{n \rightarrow \infty}\pr(\tilde{B} \text{ has a $d$-unforced path
  of length $\beta\log n$}) = 0.
$$
\end{lemma}
\begin{pf}
Since $a > c$ and $a + b + c = 1$, for sufficiently large $d$ and $n$,
there is $\alpha < 1$ such that
$$
2(1 - a + c)^d + b + 2c + o(1) \le \alpha.
$$
Then for any $\beta > 0$,
$\alpha^{\beta\log n} = n^{\beta\log\alpha}$, and by Lemma
\ref{lemunforcedpath}, the result follows.
\end{pf}
\begin{lemma}\label{lemcyclebound}
Let $a > c$. For sufficiently large $d$ and every $\epsilon > 0$,
there is $k$ such that
$$
\pr(\tilde{B} \text{ has a $d$-unforced cycle of length $\ge k$})
  < \epsilon.
$$
\end{lemma}
\begin{pf}
Take $\alpha$ and $\beta$ as in the last Lemma. Then
\begin{multline*}
\pr(\tilde{B} \text{ has a $d$-unforced cycle of length $> \beta\log n$})
\\
\begin{split}
& \le
  \pr(\tilde{B} \text{ has a $d$-unforced path of length $\beta\log n$}) \\
  & =  o(1).
\end{split}
\end{multline*}

For any $l \le \beta\log n$, by Lemma \ref{lemunforcedcycle},
\begin{align*}
\pr(\tilde{B} \text{ has a $d$-unforced cycle of length $l$})
 & \le  \alpha^l \text{ and} \\
\pr(\tilde{B} \text{ has a $d$-unforced cycle of length $\ge k$})
 & \le \sum_{l = k}^{\beta\log n} \alpha^l + o(1)
 \\
 & \le  \frac{\alpha^k}{1-\alpha} + o(1) \\
 & < \epsilon
\end{align*}
if $k > \log((1 - \alpha)\epsilon)/\log \alpha$.
\end{pf}
\begin{lemma}\label{lemnoconncycles}
Let $a > c$. For sufficiently large $d$
$$
\lim_{n \rightarrow \infty}\pr(\tilde{B}
\text{ has $d$-unforced cycles connected by a $d$-unforced
path }) =0.
$$
\end{lemma}
\begin{pf}
Let $\epsilon > 0$.
By Lemmas \ref{lempathbound} and \ref{lemcyclebound}, with probability
greater than $1 - \epsilon/2$, we can assume there are no $d$-unforced
cycles larger than $k$ and no $d$-unforced paths longer than $\beta\log n$.
Let $k_1,k_2 \le k$ and $l \le \beta\log n$. Taking $\alpha$ as in Lemma
\ref{lempathbound} and using the same argument, the probability that
there exist $d$-unforced cycles of length $k_1$ and $k_2$ connected by a
path of length $l$ is bounded above by
$2n^{-1}\alpha^{k_1 + k_2 + l}$. Summing over all
$k_1,k_2 \le k$ and $l \le \beta\log n$, the probability that there exist
$d$-unforced cycles connected by a $d$-unforced path is $\epsilon/2+o(1)$.
\end{pf}
\begin{theorem}
If $a > c$ then there is a constant $\beta$ such that
$$
\lim_{n \rightarrow \infty}
\pr(\tilde{B} \text{ has a tail longer than $\beta\log n$}) = 0.
$$
\end{theorem}
\begin{pf} We take $d$ as first used in Lemma \ref{lempathbound}.
After $d$ steps, the only gates that are not yet
stable are those in $d$-unforced paths and cycles. We may assume that all
these cycles and paths are disjoint except possibly at the endpoints of
the paths. By Lemma \ref{lemnoconncycles}, with probability $1-o(1)$,
no path begins
and ends at a cycle.

Let $l$ be the length of the longest $d$-unforced path in $\langle
\tilde{D},\tilde{\boldsymbol f} \rangle$ and $m$ be the size of its
largest $d$-unforced
cycle. After $l$ more steps, the only gates that are not yet stable are
those in $d$-unforced cycles and paths beginning at an unforced cycle.

Now consider the state of the gates in these cycles, i.e. the projection
of the state of $\langle \tilde{D},\tilde{\boldsymbol f} \rangle$,
where we look
only at the values of the gates in the $d$-unforced cycles. This state will
reach its state cycle in at most $2m$ steps. Then $\langle
\tilde{D},\tilde{\boldsymbol f} \rangle$ will reach its state cycle in at
most $l$
more steps. The Theorem then follows from Lemma \ref{lempathbound}.
\end{pf}
\begin{theorem}
If $a > c$ then for every $\epsilon > 0$ there is $s$ such that
$$
\pr(\tilde{B} \text{ has a state cycle larger than $s$}) < \epsilon.
$$
\end{theorem}
\begin{pf}
Any path (or cycle) of unstable gates is also a $d$-unforced
path (or cycle). By Lemma \ref{lemnoconncycles}, no $d$-unforced cycle is
connected by a
$d$-unforced path to a $d$-unforced cycle. Therefore no cycle of unstable
gates is connected by a path of unstable gates to a cycle of unstable
gates.
In other words, the unstable gates are partitioned into disjoint
sets, each set inducing a subgraph
of $\tilde{D}$ with one cycle and possibly some paths that begin on the
cycle but are pairwise disjoint off the cycle. Take
any input $x$, and consider the restriction $\overline{x}$ of $x$ to the
gates in any one of these partitions. Let $m$ be the size of the cycle
induced by this partition. The state cycle entered by $\overline{x}$
has size $t$ or $2t$ for some factor $t$ of $m$.
By Lemma \ref{lemcyclebound}, with probability $1 - \epsilon/2$, $m < k$.
The size of the state
cycle entered by $x$ is the least common
multiple of the sizes of the state cycles of the partitions.
This is bounded above by $s = 2\lcm(1,\dots,k-1)$.
\end{pf}
\begin{corollary}
Let $a > c$ and $\omega(n)$ be any unbounded increasing function.
Then
$$
\lim_{n \rightarrow \infty}\pr(\tilde{B} \text{ has a state cycle larger
  than $\omega(n)$}) = 0.
$$
\end{corollary}

\end{document}